\documentclass[12pt,a4paper]{article}
\usepackage{amsfonts}
\usepackage{latexsym}
\usepackage[dvips]{epsfig}
\begin{document}
\author{S.B.~Rutkevich \footnote{On leave from: Institute of Solid State and 
                Semiconductor Physics, P. Brovka str. 17, Minsk 220072, 
                Belarus.} \\
     {\it   Institute for Theoretical Physics,}\\
     {\it   University of Cologne, D-50937 K{\"o}ln, Germany}\\
     {\it   e-mail: rut@mailaps.org}} 

\title{Monte-Carlo simulation of nucleation in the two-dimensional Potts model}
\maketitle
\begin{abstract}
Nucleation in the two-dimensional $q$-state Potts 
 model has been studied by means of Monte-Carlo simulations using the 
 heat-bath dynamics. The initial metastable state has been  prepared by 
magnetic quench of  the ordered low-temperature phase. The 
magnetic field dependence of the nucleation time has been measured as the 
function of the magnetic field for different $q$ and lattice sizes at 
$T=0.5\, T_c$. A size-dependent crossover from the coalescence to nucleation 
region is 
observed  at all $q$. The magnetic field dependence of the nucleation time
is roughly described by the classical nucleation theory. Our data show increase
of the anisotropy in the shape of the critical droplets with increase of $q$.
\vspace{1cm}

\end{abstract}
\section{Introduction}
Nucleation of 'bubbles' or 'droplets' of the stable phase is a very
common phenomenon which initiates relaxation of metastable states near the
first order phase transition in many systems of condensed matter physics 
\cite{RG}, quantum fields \cite{Stan}, and cosmology \cite{Guth,Kastrup}. 
 The main quantity of interest for the
theory and applications is the nucleation time, which characterizes the time 
required by the system to leave the metastable state.
 In lack of a consistent theory of the metastability, 
the nucleation problem  has been subjected to extensive study in  
computer simulations.  Numerical data
are usually interpreted in terms of the phenomenological classical 
nucleation theory, or in terms of  
its field-theoretical version proposed by Langer \cite{Lan}. 
A great deal of this  work has been carried 
out on the Ising model supplied with some appropriate dynamics. 

It seems 
reasonable, however,  to extend the numerical study of the nucleation to 
more complex and rich models, taking into account a great variety of  physical 
realizations of this  phenomenon.   
In this paper, we have studied nucleation in the two-dimensional 
$q$-state Potts model by use of the Monte-Carlo simulation. Due to a rich 
equilibrium  phase structure, this model provides a lot of  
possibilities to form the initial  metastable phase. For 
$q>4$,  the zero-field Potts model has the first order phase transition point
driven by the temperature \cite{Bax}. Near this point, the metastable state  
can be prepared by a quenched cooling from the high to low temperature phase. 
Relaxation of such a state in different regimes has been studied numerically 
by Safran {\it et. al} \cite{Sahni1,Sahni2}, and by Arkin {\it et. al} 
\cite{Ark1,Ark2}. Meunier and Morel \cite{Meu} analysed metastability in the 
thermally quenched Potts model within the Fisher's droplet picture approach
\cite{Fish}. Arkin {\it et. al} confirmed \cite{Ark1,Ark2}, that in a case of 
slight quench (small $|T_c-T|$), the equilibrium proceeds through the usual 
nucleation mechanism. However, lack of the macroscopic 
order parameter (like the average magnetization in the Ising system below 
$T_c$) in the both initial and final states hinders precise 
measurement of the nucleation time in this regime.  
 
We simulate relaxation of the metastable states, which are 
induced in the Potts ferromagnet near the first order phase transition driven 
by the external  magnetic field $H$. The heat-bath dynamics is used to 
thermalize spins. Nucleation time is measured as the 
function of the applied field at $T=0.5\, T_c$ for $q=2,3,5,15$. 

 One can say, that for small $q$ , $q\le 4$, the line $H=0,\quad 0<T<T_c$ 
in the Potts model 
represents the liquid-vapour phase transition, and  it simulates the 
liquid-solid transition for large $q$,  $q>4$. 
It should be noted, that the Langer's theory of 
nucleation \cite{Lan} based on the coarse-grained free-energy functional can 
not be applied in the latter case \cite{Gun}.  
So, our aim is to identify in the Potts model the nucleation regimes 
well-known in the Ising model, and to study how they are modified by 
 increase of $q$.
\section{The Potts Model}
We  study the   $q$-state  ferromagnetic Potts model on the  square 
lattice 
having $L$ rows and $L$ columns. The discrete spin variable 
$\sigma =1,2,...,q$ 
is associated with each lattice site 
$j$. Helical boundary conditions are used \cite{Newman}. The model
Hamiltonian is defined as
\begin{equation}
{\cal E}=-2 J \sum_{<i\,j>} \delta (\sigma_i,\sigma_j)-H\sum_j
  \delta (\sigma_j,2).
\label{Ham}
\end{equation}
Here the first summation is over the nearest neighbour pairs, $2 J>0$ is the 
coupling constant, $H$ is the external magnetic field
applied along the $2$-direction, and 
\begin{displaymath}
 \delta (\sigma,\sigma')=\left \{ \begin{array}{ll}
 1 & \textrm{if $\sigma=\sigma' $},\\
 0 & \textrm{if $\sigma \neq \sigma' $.}
  \end{array} \right.
\end{displaymath}
We use also dimensionless parameters $K=J\beta$, and 
$h=H\beta$, where $\beta=1/k_B T$ is the inverse temperature.
The macroscopic state of the lattice can be described by $q$ real 
positive order parameters $M(\sigma)$
\begin{eqnarray}
 M(\sigma)= L^{-2} \sum_j \delta (\sigma_j,\sigma ),
  \label{m} \\
  \sum_{\sigma =1}^q  M(\sigma)=1. &
\nonumber
\end{eqnarray}
The model undergoes ferromagnetic phase transition at the critical temperature
\begin{equation}
  T_c=\frac{2 J}{k_B \log(1+\sqrt{q})}.
\end{equation}
The ferromagnetic low-temperature phase at zero field 
is $q$-times degenerated. Different ferromagnetic phases are enumerated
by the discrete parameter $\hat{\sigma}$. This implies that
 in the $\hat{\sigma}$-phase, the orientation  $\hat{\sigma}$ dominates in the 
lattice.  
For a review of many others known properties of the Potts model see \cite{Wu}.
\section{Classical Nucleation Theory}
The classical nucleation theory (CNT) considers thermally induced generation 
of the 
stable phase droplets in the metastable surrounding. A droplet of radius $R$ 
can appear with probability $p(R)$ proportional to the Boltzmann 
factor
$$
     p(R) \sim L^{2}\, \exp [-\beta {\mathcal F}(R)],
$$
where ${\mathcal F}(R)$ is the droplet free energy\footnote{In the Ising 
model, the volume free energy $-2\, \pi R^2 M\, H$ of the droplet has 
an additional factor 2 due to a different definition of the 
magnetic field $H$.}:  
\begin{equation}
 {\mathcal F}(R)=2 \pi R\, s - \pi R^2 M\, H. \label{fnuc}
\end{equation}
Here $s$ is the surface tension on the droplet boundary, $M$ is the 
zero-field spontaneous magnetization.  
The critical droplet
radious $R_c$ maximizes the free energy (\ref{fnuc}). 
It is postulated in CNT, that the nucleation 
 time   $t(h,L)$  is inversely proportional to the probability of generation 
of the critical droplet:
\begin{equation}
 t(h,L) \sim L^{-2} \exp [\beta {\mathcal F}(R_c)]=
   L^{-2} \exp [\pi(\beta s)^2/(M\, h)].   \label{istime}
\end{equation}
The classical nucleation theory considers only spherical droplets.
This is reasonable, if the surface tension of the domain wall does not depend 
on its is orientation. If the surface tension is anisotropic,  
the critical droplet are supposed to have the 
equilibrium shape given by the Wulff's construction \cite{RG}. In this 
case  equation (\ref{istime}) should be modified to 
\begin{equation}
 t(h,L) \sim 
   L^{-2} \exp[A(T)/h],    \label{antime}
\end{equation}
where the amplitude $A(T)$ is determined by the equilibrium droplet shape.  
In the Ising model, this shape is given by the equation \cite{Zia}
\begin{equation}
 \cosh x_1+\cosh x_2 =\sinh(2 K)+{1\over {\sinh (2 K)}}, \label{shape}
\end{equation}
where $x_1, \, x_2$ denote Descartes coordinates on the droplet boundary. 
It  varies from circle at $T=T_c-0$ to square at $T=0$.

The amplitude $A(T)$ can be written as $ A(T)= S(T)/M $,
where $S(T)$ denotes the area bounded by the curve (\ref{shape}),
and 
$$
M=\{ 1-[\sinh(2 K)]^{-4}\}^{1/8}.
$$   
Numerically  we have for the $q=2$ state Potts (Ising) model at $T=0.5\, T_c$:
\begin{eqnarray} 
S(0.5\, T_c)& = & 6.537, \ M(0.5\, T_c)=0.998, \nonumber  \\
 A(0.5\, T_c)& = & 6.55. \label{val}
\end{eqnarray}
 It should be noted, that equation (\ref{antime}) gives only the 
leading $h \to 0$ asymptotics of the nucleation time. The most important 
correction is expected to have the form of the prefactor with a power-law 
 $h$-dependence \cite{RG}
 \begin{equation}
 t(h,L) = 
   L^{-2} \frac{B(T)}{h^{b+c}}\exp[A(T)/h].    \label{tcor} 
\end{equation}
The exponent $b$ here arises in the field theoretical nucleation theory from 
the contribution of the capillary waves on the critical droplet 
surface \cite{W}, $b=1$ in 
the two-dimensional Ising model. The exponent $c$ gives the $h$ variation of 
the so-called kinetic prefactor, which depends on the chosen dynamics. For 
dynamics which can be described by the Fokker-Plank equation, the value 
$c=2$ is expected \cite{Lan2,Lan3}.

The exact equilibrium (Wulff's) shape  of the droplet is not known 
 for the Potts model with $q\ge 3$. So, we had to use more simple and less 
accurate estimates for the amplitude $A(T)$. We have 
used  square and circle approximations for the critical droplet shape.
The surface tension in the both cases was replaced by its zero-temperature 
value $s =2J$. This yields for the amplitude $A(T)$ in the nucleation time 
exponent:
\begin{eqnarray}
 A(T)=\left \{ \begin{array}{ll}
 16 \, K^2,  & \textrm{for square droplet}, \label{A}\\
 4 \pi \, K^2. & \textrm{for circle droplet}
  \end{array} \right.
\end{eqnarray}

Numerical values for the amplitude $A(T)$ calculated from
(\ref{A}) for different  $q$ at $T=0.5\, T_c$ are given in Table \ref{tA}. 
\begin{table}[hbt]
 \caption{Amplitude $A(T)$ at $T=0.5\, T_c$ for square and circle droplets.} 
  \label{tA}
 \begin{center}
  \begin{tabular}{|c|c|c|}\hline
  q& \multicolumn{2}{|c|}{$A(0.5\, T_c)$}\\ \hline

     & square       & circle  \\ \hline
   3 & 16.2         & 12.7  \\ \hline
   5 & 22.1         & 17.3 \\ \hline
   15& 40.1         & 31.5 \\ \hline 
  \end{tabular}
 \end{center}
\end{table}

The size-dependent behaviour (\ref{antime}) of the nucleation time is 
observed in not very large
lattices, in the so-called nucleation regime. In this regime, only one
droplet of the stable phase appears in the system, and then expands to 
the whole lattice. 
 
At large $L$, the coalescence  regime is
realized. In this regime the relaxation time does not depend on $L$. Many 
critical droplets appear in the system, expand and coalesce.  
 
It is expected from the Kolmogorow-Johnson-Mehl-Avrami theory \cite{Sek}, that 
\begin{equation}
  t(H) \sim  \exp [ A(T)/(3 h)] \label{coa}
\end{equation}
in the coalescence regime in  two-dimensional systems .
\section{Simulation Scheme}
In the Potts model, the applied field and the initial metastable configuration
of spins in the lattice can be chosen in many different ways. We have 
studied only two of them.
\begin{itemize}
\item {\bf First regime}.  A positive magnetic field $H>0$ is applied along 
the $2$-direction, 
whereas all the spins are initially oriented along the $1$-directions:
 $\sigma_j = 1$. Such a field removes completely degeneracy of the 
low temperature phase, leaving only one  equilibrium ferromagnetic state: 
$\hat{\sigma}=2$.
\item {\bf Second regime}. A negative magnetic field $H<0$ is applied 
along 2-direction, and all 
spins are  initially oriented in the same direction  $\sigma_j=2$.  The 
ferromagnetic phase still remains $(q-1)$-times degenerated (if $q \ge 3$) in 
such a field. In this regime, the system moves to some bifurcation point, 
and then relaxes to one of the equivalent $(q-1)$ equilibrium states.   
\end{itemize}
The  heat-bath technique (Glauber kinetics) has been used to orient
the spins. To produce a sequence of pseudorandom numbers, we have used the 
standard generator of pseudorandom numbers with multiplication by 16807 to 
give 64-bit integers. 
 The candidate spins to flip have been chosen sequentially: 
site by site in the row, and row by row in the lattice.
We define the nucleation time $t$ as the number of sweeps through the 
lattice after which the number of spins in the initial (metastable) phase 
reduces to $ 0.5 L^2$. The $h$-dependence of the nucleation time $t$ has 
been measured at $L=31$, and $L=1001$. In the nucleation regime, the 
nucleation time fluctuates very much. So, we had to average the nucleation 
time in this regime over 100 or even 1000 runs. 
It is important to note, that for $q\ge 3$ the relation between the nucleation 
time $t$ and the total relaxation time $t_0$ is not straightforward 
and is  different in different  regimes.

A typical time dependence of the order parameters $M(\sigma)$ in the first
nucleation regime for $q=3$ is shown in Fig. 1. Since the fall of 
the magnetization $M(1)$ is very rapid compared with the time which the 
system has spent in the metastable state, the nucleation time $t$ and the 
total relaxation time $t_0$ are essentially the same in this regime. This
is true also in the second nucleation regime. In the latter case, however, 
the system can fall with equal probabilities to one of the 
equilibrium states, either  $\hat{\sigma}=1$,  or $\hat{\sigma}=3$.

Fig. 2 shows a typical time evolution of the magnetization in the 
first coalescence regime. Up to the saturation at 1, the magnetization $M(2)$ 
increases approximately linearly with time due to the steady expansion of
 many stable phase droplets. Thus, the system reaches equilibrium in about 
two nucleation times: $t_0 \simeq 2\, t$.

The situation is quite different in the second coalescence regime, see 
Fig. 3. In this case many critical droplets of the both equilibrium
phases $\hat{\sigma}=0$ and $\hat{\sigma}=1$ are initially generated in the 
lattice. They expand and expel the metastable $\hat{\sigma}=2$ phase from the 
system in about two nucleation times. The resulting state is still far from the
equilibrium, since the lattice is divided into domains of two stable 
phases $\hat{\sigma}=1$ and $\hat{\sigma}=3$. During the final (very slow at 
low temperatures) stage of the relaxation, these two phases compete, until one
of them will expel another from the system. So, the nucleation time $t$ and 
 the total relaxation time $t_0$ may be of different order of magnitude in 
the second coalescence regime. 
\section{Results}
Our simulations were restricted to the temperature $T=0.5\, T_c$ 
and the lattice
sizes $L=31$ and $L=1001$. We have started from the $q=2$ Potts model, which 
is equivalent to the Ising model. This provides a good possibility to verify 
efficiency of the applied simulation scheme, since nucleation in the Ising 
model has been thoroughly studied \cite{RG,St1,RT,Wo,St2}.

Figure 4 shows the nucleation time in the logscale for the $q=2$ Potts
(Ising) model. Pluses correspond to the nucleation time averaged over 
1000 runs in the $L=31$ lattice. Stars relate to the nucleation time in 
the $L=1001$ lattice. Crosses indicate the dispersion $D[t]$ of the 
 nucleation time  distribution in the $L=31$ lattice:
$$
 D[t]=(\langle t^2\rangle -  \langle t\rangle^2)^{1/2},
$$
where $\langle ...\rangle$ denotes averaging over different runs. 

Three relaxation regimes are clearly identified: the strong field regime 
(SFR), the coalescence regime (CR), and the nucleation regime (NR). Increase 
of the lattice size does not change the nucleation time in the 
strong field and coalescence regimes. The nucleation
regime was not observed in the $L=1001$ lattice in the studied 
region of fields $h$. In the nucleation regime, the nucleation time fluctuates
very much, and the distribution $p(t)$  of these fluctuations is described
by the simple exponential law: 
\begin{equation}
 p(t)={1\over \tau} \exp(-t/\tau).  \label{texp}
\end{equation}
For this distribution, the averaged time equals to the dispersion:
$ \langle t \rangle = D[t]= \tau $.
This is indeed observed in the $L=31$ lattice at $1/h>1.15$ indicating
well established nucleation regime in this field region. 
The theoretical prediction for the averaged nucleation time in the nucleation 
regime with the Wulff's shape droplet is given by equation (\ref{tcor}), 
where  $b+c=3$, and $A=6.55$.
The upper straight line in Fig. 4  shows the nucleation theory
exponent, and the second (curved) line plots the the same exponent 
corrected
by the prefactor $C h^{-3}$, i.e. equation (\ref{tcor}). Though the latter 
curve contains only one
free parameter - the constant factor $C$ -  it fits well the 
numerical simulation points
for the $L=31$ lattice in the nucleation regime, much better then 
the first (straight) line.  

The curved line gives also a good fit for the 
dispersion points in the small $L=31$ lattice not only in the nucleation 
region, but also at crossover to the coalescence regime down 
to $h^{-1}\approx 0.8$. The possible explanation of this result may 
be the following. Measured nucleation time $t$ consists of two 
terms $t=t_n+t_e$, which have the same order of magnitude in the crossover 
region. The first term relates to the time of nucleation of the first critical 
droplet in the system, and the second term 
denotes the time of the expansion of this droplet to $1/2$ of the whole 
lattice. The first term should strongly fluctuate according to the 
exponential law (\ref{texp}). If the second term $t_e$ fluctuates much weaker,
one  should have for the total time dispersion:
$$
 D[t]=D[t_n+t_e] \approx D[t_n]=\langle t_n \rangle =\tau,
$$
where the averaged time  $\tau$ of the first droplet nucleation is still given
 by equation 
(\ref{tcor}). Thus, in the crossover region, dependence (\ref{tcor}) should 
fit better the time dispersion points $D[t]$, than the averaged time 
$\langle t \rangle $, which is indeed observed. 
 
The first numerical verification of the magnetic field dependence of the 
prefactor in 
(\ref{tcor}) for the two-dimensional Ising model  has been reported by 
Rikvold {\it et. al} \cite{RT}. In this work, the value $b+c=3$ was confirmed  
at $T=0.8\, T_c$ for the Metropolis dynamics with randomly chosen sites. 
However, a smaller value $b+c \approx 2$ was observed in \cite{RT}, when the 
sites in the lattice were chosen sequentially. Though we have also used the 
sequential chose of sites in the lattice, our data for the nucleation time in 
the nucleation region are better fitted 
by (\ref{tcor}) with the prefactor exponent  $b+c=3$, rather then $b+c=2$, see 
Fig. 5. This is more clear for the dispersion $D[t]$ (crosses in 
Fig. 5), since the latter is still described by (\ref{tcor}) at 
the crossover to the coalescence regime.  

The lower straight line is the best linear fit for the nucleation time in the
larger $L=1001$ lattice in the coalescence regime (CR). The ratio of 
slopes $R=1.9$
of two straight lines in Fig. 4  corresponding to the nucleation and 
coalescence regimes is much smaller than $3$, expected from the theory in 
the small $h$ limit. However,  we  are still far from the well-established 
small $h$ limit, as it is evident  from the considerable difference between 
first and second lines in Fig. 4. If one takes this into account,
this disagreement of the observed 
$R=1.9$ value with $3$  expected in the $h\to 0$ limit  becomes not so 
surprising.

Figures 6-10  show the observed dependence of the nucleation 
time on the 
inverse field $h^{-1}$ in the $q=3,5,15$ Potts models, $L=31,1001$ in 
the first 
 $h>0$ and second $h<0$ regimes. Pluses correspond to the nucleation time 
averaged over 100 runs  in the smaller lattice. Crosses indicate nucleation 
time in the $L=1001$ lattice. Only the second  regime $h<0$ has been 
 studied in the $q=15$ Potts model, since the nucleation time in the first
regime $h>0$ becomes too long in this case, see equation (\ref{tratio}) 
below. 

Qualitatively, the  magnetic field dependence of the nucleation 
time in  all cases 
is similar to that in the Ising model. The same three field regions 
(SRF, NR, CR) have been observed. As in the Ising case, the nucleation regime 
was realized only in the 
smaller lattice, $L=31$. In the latter regime, the relaxation time strongly 
fluctuates. 

In all cases, $\log \langle t\rangle $ behaves linearly with $1/h$ in the 
nucleation regime (NR), in agreement with the classical nucleation theory. 
In the $q=3$ and $q=5$ systems the observed slopes 
($\log \langle t\rangle\,{\textrm vs.}\,  h^{-1}$ ) 
in NR agree with the circle shape of the critical droplets. However,  in the 
$q=15$ model our data indicate the square critical droplet shape, 
see Fig. 10 , and Table \ref{tA}.

The ratio of slopes in nucleation and coalescence regimes varies from
$R=2.6$ in the $q=3$, $h>0$ case, down  to $R=2.1$ in the $q=5$, $h<0$ case.
As in the $q=2$  model, these values are lower than $3$ predicted by the 
theory. 

In the first nucleation regime ($h>0$), the nucleating critical droplet should 
be of the single equilibrium phase $\hat{\sigma}=2$, whereas in the second 
nucleation regime ($h<0$),  a critical droplet of either of $(q-1)$ 
equilibrium phases $\hat{\sigma}, \  \hat{\sigma}\ne 2$ provides relaxation.
 So, it is naturally to expect, that nucleation time $\langle t_1 \rangle $ in
 the first 
nucleation regime will be longer, that the nucleation time 
$\langle t_2 \rangle $ in the 
second nucleation regime, and 
\begin{equation}
  {\langle t_1\rangle \over \langle t_2\rangle } \approx q-1.  \label{tratio} 
\end{equation} 
The observed ratios $\langle t_1 \rangle /\langle t_2 \rangle$  determined 
from the circle droplet fits in Fig. 6-9
\begin{eqnarray}
  {\langle t_1\rangle \over \langle t_2\rangle }\approx    
  \left \{ \begin{array}{ll}
   \exp(-2.95)/ \exp(-3.64)=2.0  & \textrm{for $q=3$}, \label{fq}\\
   \exp(-1.29)/ \exp(-2.8)=4.5 & \textrm{for $q=5$}  
  \end{array} \right.
\end{eqnarray}
are in a good agreement with (\ref{tratio}).
\section{Summary} 
We have simulated  nucleation in the $q$-state Potts model 
on the square lattice for $q=2,3,5,15$.
The magnetic field dependence of the nucleation time have been measured in 
two regimes ($h>0$ and $h<0$) at $T=0.5 \, T_c$  for two 
lattice sizes $L=31$ and $L=1001$.  The strong field regime, coalescence 
regime, and nucleation regime are clearly identified. 

The observed magnetic field dependencies of the nucleation time 
is in agreement with the classical nucleation theory, if one takes into 
account the anisotropy of the critical droplet shape.   
Our numerical data indicate, that the shape of the critical droplet changes 
from (roughly) circle towards square with increasing $q$ at the fixed 
temperature $T=0.5\, T_c$. A similar result was observed by Bikker {\it et. al}
 \cite{Bar} in numerical calculations of the equilibrium crystal shape in the 
three-dimensional Potts model. So, increase of $q$ has a similar effect on 
the equilibrium droplet shape, as decrease of temperature. This is natural, 
since the both reduce the correlation length, which should increase 
anisotropy in the system.

The ratio of 
($\log (t) \,{\textrm vs.}\,  h^{-1}$) slopes in nucleation and coalescence regimes does not 
approach 3, predicted by the theory in the $h \to 0 $ limit. 
We associate this discrepancy with the 'finite-$h$ corrections', which are 
still considerable in the studied region of the magnetic field. 

Though 
$q=2,3$ and $q=5,15$ state Potts models have different type phase transitions 
at $T_c$ (continuous for $q=2,3$ and first order for $q=5,15$), metastable 
phases in these models relax to the equilibrium in a similar way.

%
\vspace{.5cm}
\noindent
{\large \textbf{Acknowledgments}}
\vspace{0.3cm}

\noindent
I am grateful to Kurt Binder for suggesting the subject of 
this work. I am thankful to Dietrich Stauffer for many helpful 
discussions, and for introducing into computer simulations.

\noindent
This work is supported by the Deutscher Akademischer Austauschdienst (DAAD)
and by the Fund of Fundamental Investigations of Republic of Belarus.

\newpage
\begin{figure}[hbt] 
\input{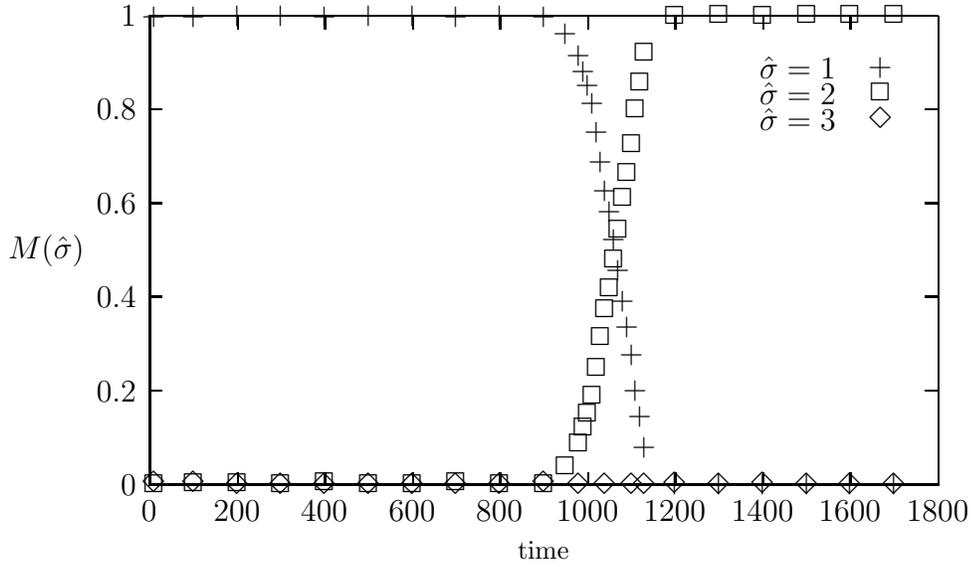}
\caption{Decay of metastable state in first nucleation regime}\label{1nu}
\end{figure}
\begin{figure}[hbt] 
\input{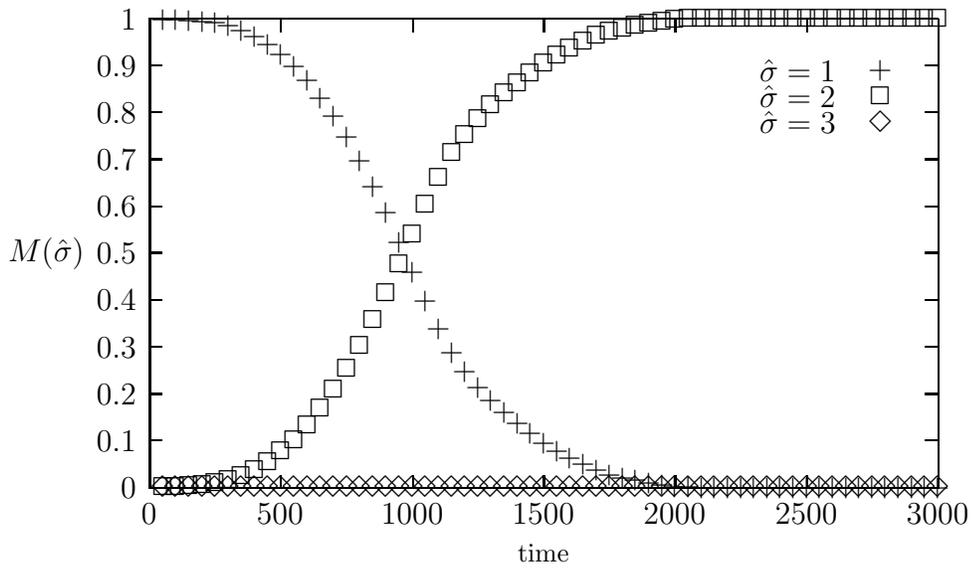}
\caption{Decay of metastable state in first coalescence regime} \label{1co}
\end{figure}
\begin{figure}[b]
\input{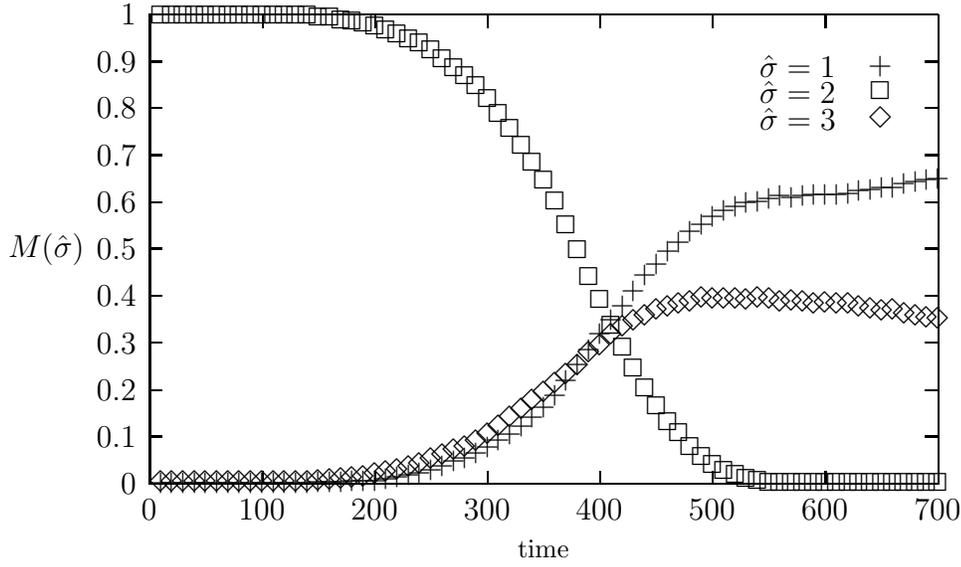}
\caption{Decay of metastable state in second coalescence regime}\label{2co}
\end{figure}
\begin{figure}[hbt]
\vspace{.8cm}
\centering
\epsfig{file=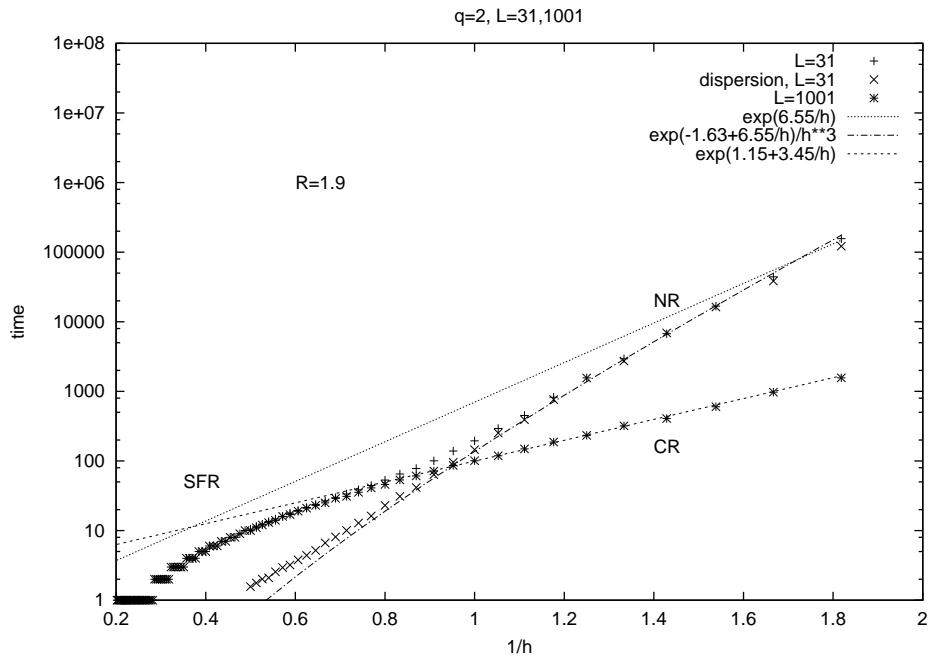}
\parbox[t]{0.8\textwidth}{
\caption{Nucleation time and dispersion plotted against $1/h$ in the $q=2$ 
   Potts (Ising)
  model for $L=31$} 
 \label{q2}}
\end{figure}
\begin{figure}[hbt]
\vspace{.8cm}
\centering
\epsfig{file=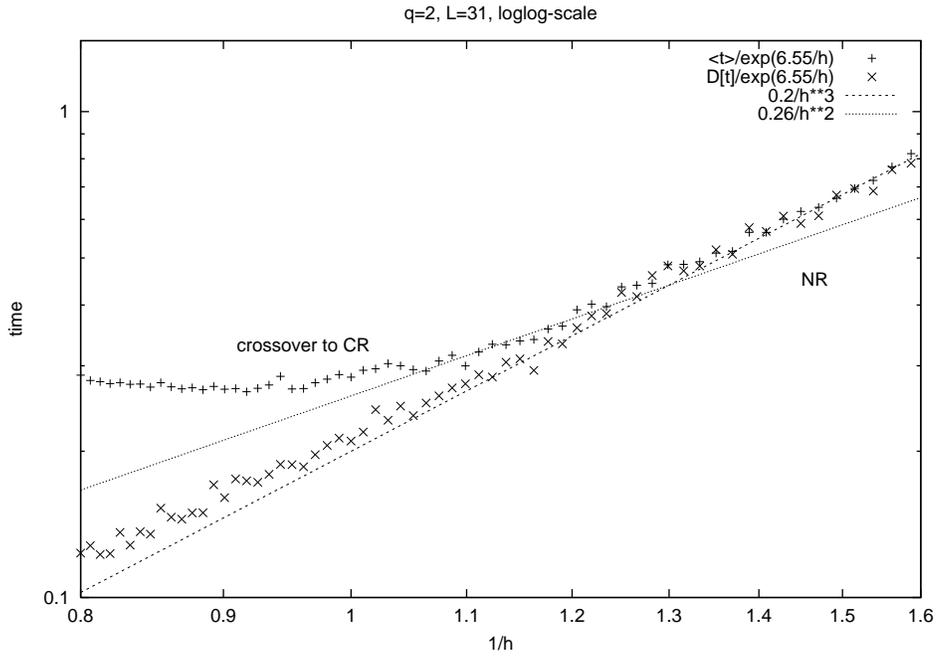}
\parbox[t]{0.8\textwidth}{
\caption{Averaged nucleation time $\langle t \rangle $ and  dispersion 
  $D[t]$ divided by $\exp[6.55/h]$,  versus $1/h$ (loglog-scale) 
  in the $q=2$ Potts (Ising) model.} 
 \label{pref}}
\end{figure}
\begin{figure}[hbt]
\vspace{.8cm}
\centering
\epsfig{file=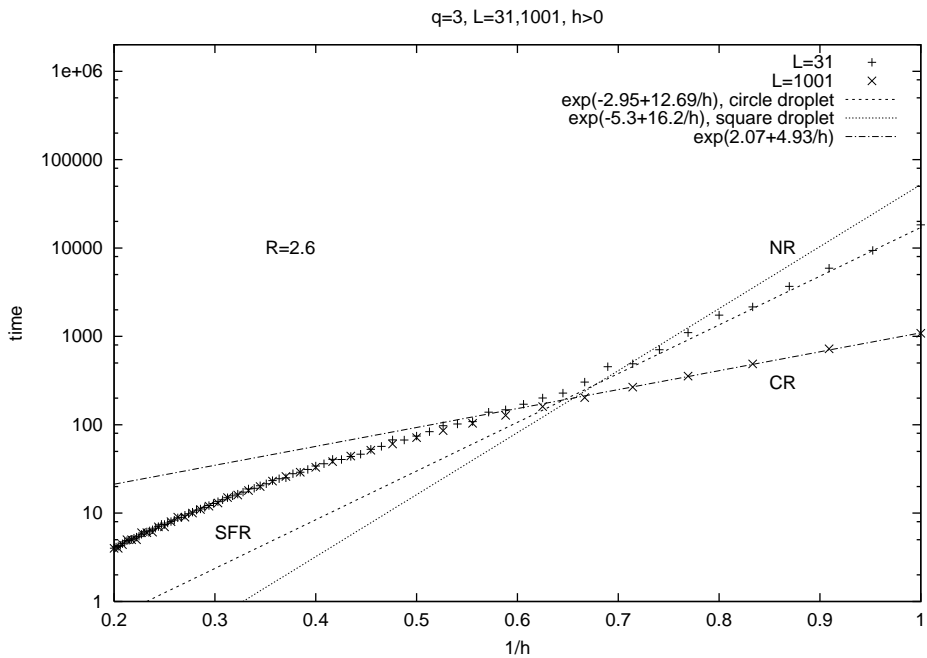}
\parbox[t]{0.8\textwidth}{
\caption{Nucleation time versus $1/h$ in the $q=3$ Potts
  model, $h>0$.\label{q3}} 
 }
\end{figure}
\begin{figure}[htb]
\vspace{.8cm}
\centering
\epsfig{file=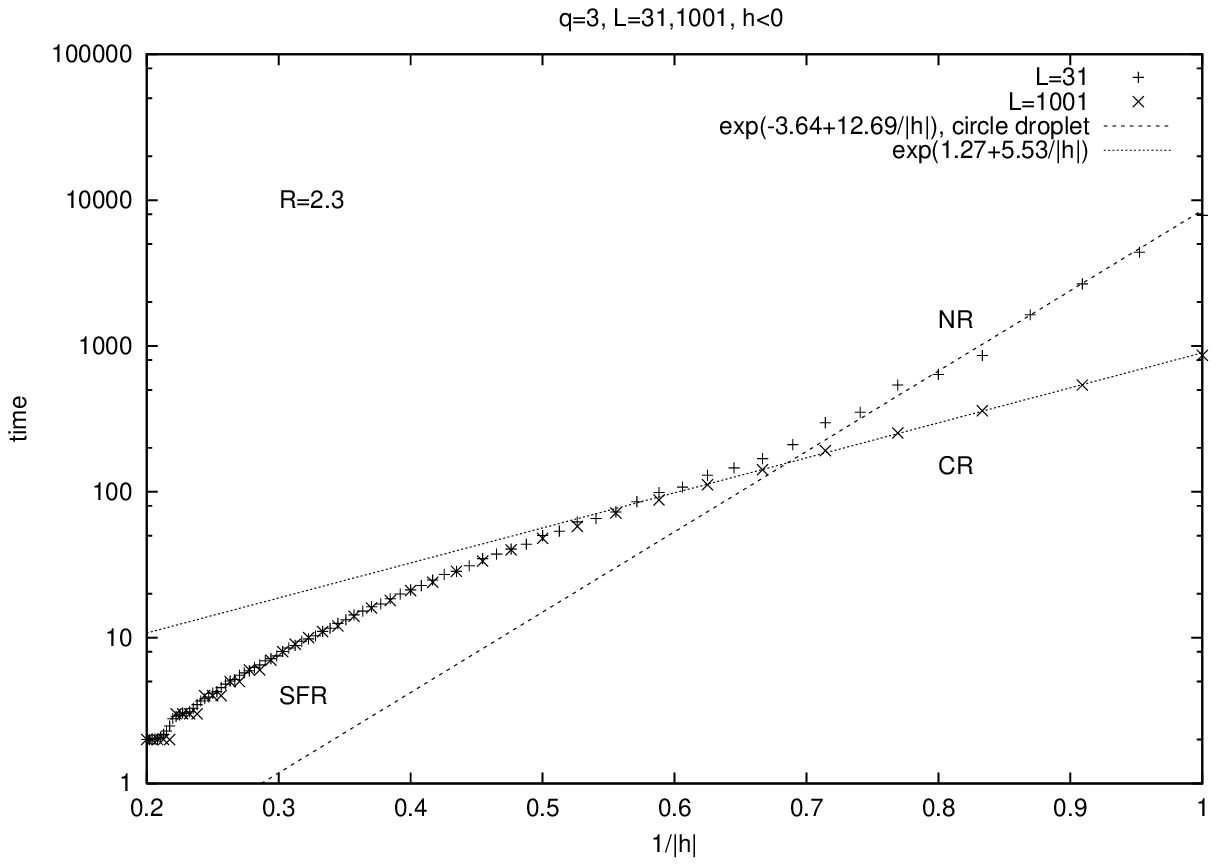}
\parbox[t]{0.8\textwidth}{
\caption{Nucleation time plotted against $1/|h|$ in the $q=3$ Potts
  model, $h<0$.\label{q3neg}} 
 }
\end{figure}
\begin{figure}[htb]
\vspace{.8cm}
\centering
\epsfig{file=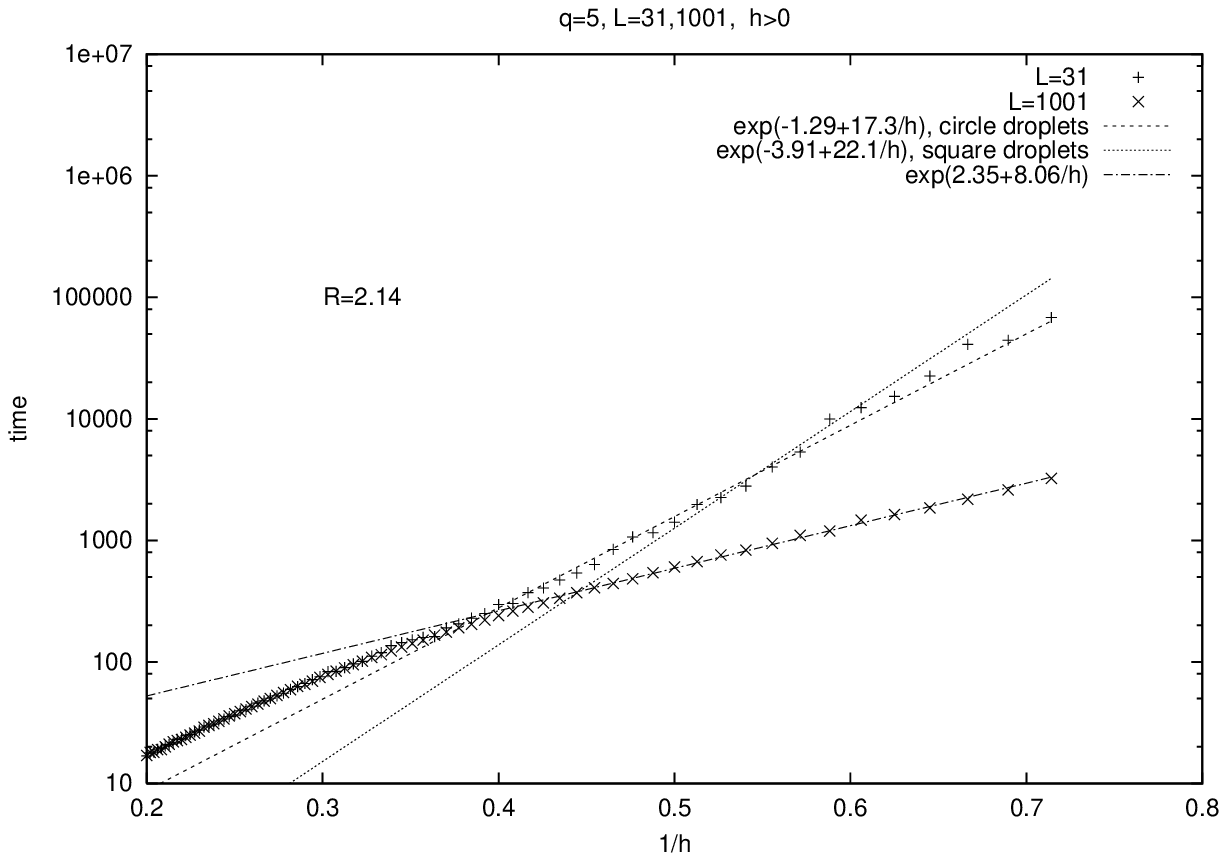}
\parbox[t]{0.8\textwidth}{
\caption{Nucleation time plotted against $1/h$ in the $q=5$ Potts
  model, $h>0$.\label{q5pos}} 
 }
\end{figure}
\begin{figure}[htb]
 \vspace{.8cm}
 \centering
 \epsfig{file=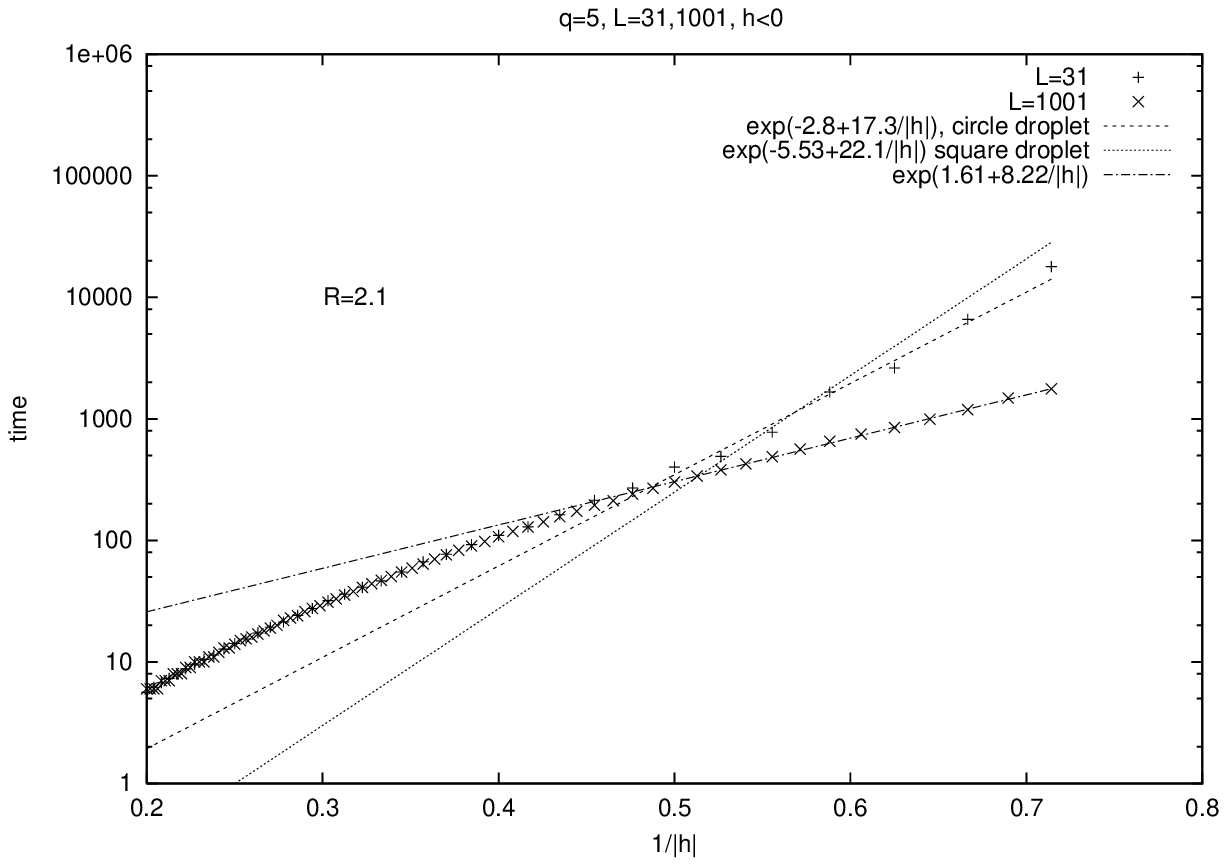}
 \parbox[t]{0.8\textwidth}{
 \caption{Nucleation time plotted against $1/|h|$ in the $q=5$ Potts
  model, $h<0$.\label{q5neg}} 
 }
\end{figure}
\begin{figure}[hbt]
 \vspace{.8cm}
 \centering
 \epsfig{file=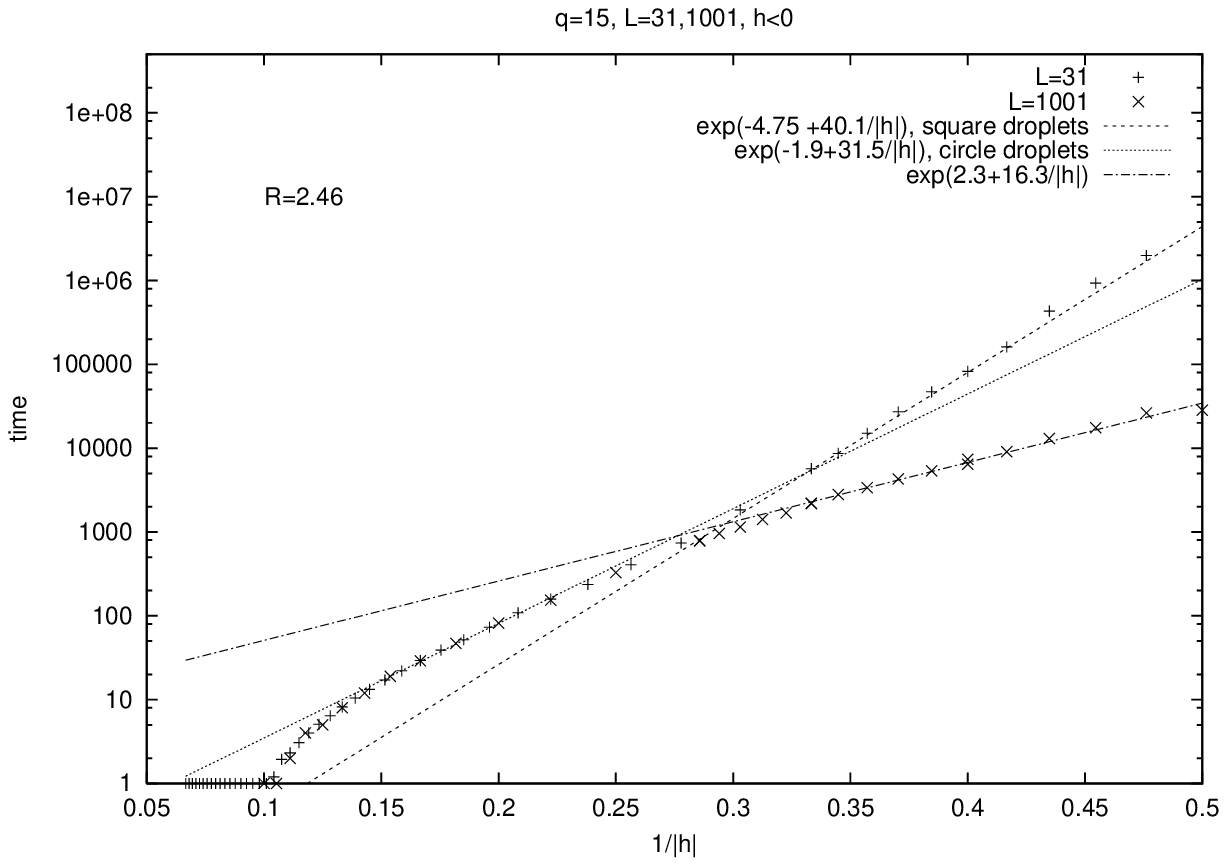}
 \parbox[t]{0.8\textwidth}{
 \caption{Nucleation time plotted against $1/|h|$ in the $q=15$ Potts
  model, $h<0$.\label{q15}} 
 }
\end{figure}
\end{document}